\documentclass[aps,prl,superscriptaddress,showpacs,floatfix,twocolumn]{revtex4}

\usepackage{graphicx}

\begin{document}

\title{Empirical estimation the effects of flow on thermal photon angular distribution and spectra in nucleus-nucleus collisions at RHIC and LHC }
\author{V.~S.~Pantuev\\
Institute for Nuclear Research, Russian Academy of Sciences, Moscow, Russia}

\date{\today}

\begin{abstract}
Experimental data for hadron radial and elliptic flow are used  
 to investigate their influence on  the shape of the thermal photon
 spectrum at RHIC. Leaving alone
the actual mechanism of photon production and its time evolution, we
concentrate on the spectrum shape. 
 Radial and longitudinal flow  of the bulk 
can  change significantly the observed photon energy spectrum via
the Doppler Effect. 
Experimental thermal photon data are described by
local frame temperature parameter which depends on the assumption of
longitudinal and radial flow. 
From the
observed hadron elliptic flow we estimate the modulation of radial
flow parameter 
 versus the angle relative to the reaction plane. Based on this 
we calculate elliptic flow parameter for thermal photons, which was found to be
very close to that for hadrons.  Considering very similar
amplitude for hadron elliptic flow at LHC and RHIC, we demonstrate that
thermal photons at LHC should show large elliptic flow as
well. All of these considerations are also valid   
  for the low invariant mass dilepton pairs.
\end{abstract}



\maketitle

Measurements of the photons radiated in nucleus-nucleus collisions are of great current interest. This interest stems from good
photon penetration through strongly interacting matter. 
From a theoretical point of view, the way to study thermal photons would be to perform a full hydrodynamic
calculation. Nevertheless, such calculations use many assumptions and parameters 
such as the initial energy and density profile,
equation of state, treatment of exact photon production channels,
initial temperature profile, equilibration time, consideration
of the quark-gluon and hadron phases and many others; these assumptions can serve to mask
many of the general features of photon emission. One of the common
effect on top of all these assumptions is the relativistic radial and 
longitudinal expansion of
the system. Such expansions can induce a Doppler Effect with red and blue shifts
for radiated thermal photons. Here, we investigate the influence of such effects on the emission spectrum for thermal photons. We do not pretend
to calculate the exact photon yield nor specify the actual mechanisms for photon production.

One can distinguish two
types of photons: 
prompt or direct photons, produced by primary partons 
in the initial stage of a 
collision, and thermal photons 
radiated by hot and dense matter after the system thermalizes. We limit
ourselves to consideration of the latter type. We start from
a simple system with fixed temperature, that expands in the transverse and longitudinal directions. Then, we examine how
additional assumptions can change this simple picture.

 The non-relativistic Doppler Effect for a frequency change from  
a  moving source, say, towards an observer is $\omega=\omega_0/(1-v/c)$, where $\omega_0$ 
is the original frequency in the rest frame of the source, $v$ is source velocity and 
$c$ is the speed of the signal in the medium. For the relativistic case, this formula changes to: 
\begin{equation}
\omega = \omega_0\frac{\sqrt{1-\beta^2}}{1-\beta cos\theta}.
\label{eq:relat}
\end{equation}
Here $\beta$ is speed of the source relative to the speed of light and $cos\theta$ is
the angle between the direction of the moving source and the observer. One can see an additional factor 
$\sqrt{1-\beta^2}$ for the relativistic case. 

For non-relativistic case, 
if the radiation spectrum has a simple Boltzman exponential shape with temperature $T$,
\begin{equation}
dN/d\omega_0=exp(-\omega_0/T) ,
\label{nonrelat}
\end{equation}
using relations $\omega_0=\omega(1-\beta)$ and $d\omega_0/d\omega=1-\beta$, we get 
radiation spectrum seen in the laboratory frame: 
\begin{equation}
dN/d\omega=\frac{dN}{d\omega_0}\frac{d\omega_0}{d\omega}=(1-\beta)exp(-\frac{\omega}{T/(1-\beta)}).
\label{nonrelat_sp}
\end{equation}
For relativistic kinematics and photons with energy $E=\hbar\omega$ the last formula changes to:
\begin{equation}
dN/dE=\frac{1-\beta cos\theta}{\sqrt{1-\beta^2}}exp(-\frac{E(1-\beta cos\theta)}{T \sqrt{1-\beta^2}}).
\label{relat_sp}
\end{equation}

To separate longitudinal and transverse expansion in nucleus-nucleus collisions, we describe a boos in the 
 transverse direction by velocity $\beta_T$ or rapidity $y_T$, and 
expansion in longitudinal direction by
velocity $\beta_L$ and rapidity $y_L$. The rapidity and velocity are related 
by the formula $\beta=tanh(y)$, 
where $tanh$ is 
a hyperbolic tangent. The total rapidity $y$ can be calculated 
 via Pifagora's relation in hyperbolic rapidity space: $cosh(y)=cosh(y_T)\cdot cosh(y_L)$. 
For an observer at mid-rapidity, radial expansion of the created matter will lead to a blue shift 
of the spectrum when matter moves exactly towards the observer. In contrast it will induce a red 
shift for photons 
produced from the medium, which expands in opposite to the observer
direction.  This expansion in a direction opposite to the observer is very similar to astrophysical expansion of our Universe. 
For different  
cells of the medium the vectors for radial flow are oriented at some angle to the
observer, this effect will be controlled by the $\beta_T cos\theta$ component 
of the flow. Thus we can write:

\begin{equation}
dN/dE=\frac{1-\beta_T cos\theta}{\sqrt{1-\beta^2}}exp(-\frac{E(1-\beta_T cos\theta)}{T \sqrt{1-\beta^2}}).
\label{eq:relat_sp}
\end{equation}

To make quantitive estimates for nucleus-nucleus collisions we run a simple simulation.
We use a density profiles of wounded nucleons for Au+Au 
collisions in $x-y$ plane transverse to the beam direction from work~\cite{jjia}, which were also used later in paper~\cite{formtime}. These 
profiles were generated for different centralities assuming Woods--Saxon density distribution of 
the colliding nuclei. 
We use a linear transverse 
velocity profile for radial flow which rises from zero at the center of the collision zone to a maximum value 
of $\beta_{T}$=0.75$c$~\cite{ppg009}. This value describes well the momentum spectra of identified 
secondary hadrons produced in Au+Au collisions at RHIC. The main results of our calculation do 
not change much by changing $\beta_{T}$ in the range 0.7--0.8$c$. 
Then, each point in $x-y$ plane was taken with the weight of relevant participant 
 nucleon density. For simplicity we consider the most central 0-5\% events and set the observer in 
$x$ direction in reaction plane of the collision. For each point we estimate the distance from the 
collision center, calculate the radial boost $\beta(r)=\beta_T \cdot r/R$  and its projection to the 
observer direction   
$\beta_T \cdot cos\theta$, where the angle $\theta$ in our set is the angle relative to the reaction 
plane 
(usually marked as $\phi$). To compare with experimental data we have
to use the invariant yield by  
multiplying Eq.~\ref{eq:relat_sp} by factor $E/p^2=1/E$ for massless photons. 

\begin{figure}[thb]
\includegraphics[width=1.0\linewidth]{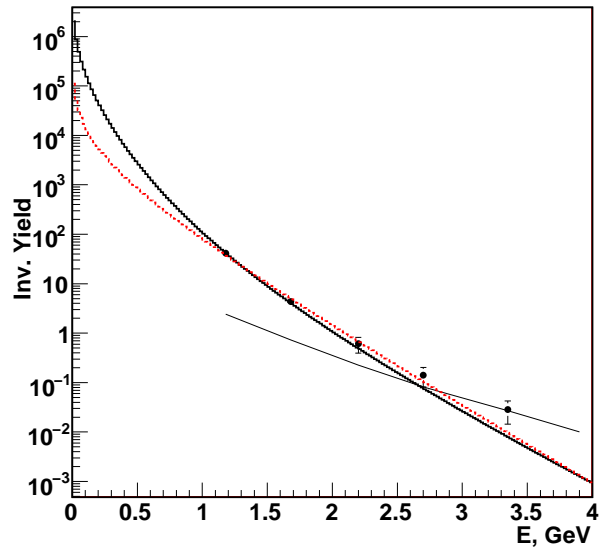}
\caption{\label{fig:data}Fig.1. Invariant thermal photon yield for central 0--5\% Au+Au collisions versus the observed photon 
energy in laboratory frame for the original source temperature T=0.17 GeV, $\beta_T$=0.75$c$ and 
$\beta_L$=0.5 -- thick solid line. The yield was normalized to the first experimental point at 1.2 GeV. 
Points are experimental data for central 0--20\% Au+Au
collisions from~\cite{data}. Thin solid line is for the experimental data for p+p collisions 
from the same measurement scaled by the mean number of binary nucleon-nucleon collisions at
this centrality. Dashed line is an estimate of photon yield at very early times of the 
collision just after thermalization occure with temperature 0.27 GeV but with no flow.}
\end{figure}

In Fig.~\ref{fig:data} we present results of the calculation. The curves were normalized to 
the first 
experimental point at 1.2 GeV~\cite{data}. We fit data for $\beta_L$=0.5. 
In this case the shape of the experimental spectrum 
could well 
be described at temperature $T$=0.17 GeV. There is a significant blue shift of the spectrum at high energy and observable red shift at low energy.  
From the formulas presented above, one can see that there 
is a strong 
correlation between parameters $T$ and $\beta_L$. 
Thus, for $\beta_{T}$=0.75$c$ the same result could 
be obtained for different parameter combinations, like $T$=0.2 GeV, $\beta_L$=0.7$c$ and
$T$=0.3 GeV, $\beta_L$=0.88$c$ or 
with no longitudinal flow when $\beta_L$=0$c$, we get  
$T$=0.15 GeV. The experimental data and our 
estimation for $E$ below 2.5 $GeV$ are very close.
Deviations from experimental points at higher  
energy could be explained by a significant contribution from direct photons, whose yield 
could be estimated by using   
 p+p data from~\cite{data} and scaled by the number of binary nucleon-nucleon collisions, 
thin solid line in figure. The yield of  
direct photons becomes dominant for photon energy above 2.7 $GeV$.  In addition, we check how our results
may change 
by the varying transverse velocity profile (from liner to
quadratic) and by introducing a temperature profile (more hot in the center and
colder at the edges). We did not find a significant change of the results with these assumptions.

In nucleus-nucleus collisions the produced matter 
definitely expands in the longitudinal direction and $\beta_L>0$, so T=0.15 $GeV$ is the lowest 
estimate of
the mean possible temperature of the system. At the same time, it is hard to get the actual temperature 
from the experimental 
spectrum without knowing the value of the longitudinal expansion. 

Another test of our model assumptions is the consideration of elliptic flow for thermal photons. 
We can make an empirical estimation of the elliptic flow parameter for
thermal photons using experimental data for hadron elliptic flow.  
For hadrons produced 
in A+A collisions we can attribute the observed azimuthal anisotropy to an amplitude modulation of the radial flow versus
the reaction plane. Elliptic flow is usually described by the parameter  $v_2$ as: 
$\sigma(\phi)=const\cdot(1+2\cdot v_{2}cos(2\phi))$, 
where $\phi$ is the angle with respect to the reaction plane.
For hadrons,the azimuthal parameter
$v_2$,  scales with number of constituent quarks and 
transverse kinetic energy per quark~\cite{ppg062}.  The value of $v_2$
per quark saturates at about 
$v_2$=0.07 at kinetic energy around 1 $GeV$ per quark for
mid-central collisions. 
We use this to estimate the difference 
of radial flow $in$ and $out$ of the reaction plane, $\beta^{in}$ and $\beta^{out}$, to get a value of $v_2$. 
The parameter $v_2$ by itself defines the relative variation of the production cross section 
$\sigma$ with
respect to the reaction plane orientation. 
So, the difference from the average yield for 
in-plane or out-plane is  
$\Delta \sigma /\sigma =2\cdot v_{2}$.

Knowing the shape of the cross
section spectrum versus the kinetic energy we can estimate what should
be the difference in kinetic energy to get such cross
section change. In the kinetic energy range under consideration
the cross section has an exponential shape with the inverse slope parameter
$T$ around 200 $MeV$ for pions and 300--330 $MeV$ for
protons~\cite{ppg009,ppg026}. Thus, for the spectrum of
constituent quarks the inverse slope parameter $T_q$ would be
around 100 $MeV$.  
If the cross section is described by a simple exponent, 
$\sigma=const\cdot exp(-E_{kin}/T_q)$, then its relative variation for an 
energy change $\Delta E_{kin}$ will be
$\Delta \sigma/\sigma=\Delta E_{kin}/T_q$ . 
To satisfy these changes we get the necessary relation with $v_2$:
\begin{equation}
\Delta E_{kin}=2\cdot T_q \cdot v_{2}. 
\label{eq:ss2}
\end{equation}

If a quark with mass $m_q$ moves with velocity $\beta$ and $\gamma=1/\sqrt{1-\beta^2}$,
then its kinetic energy $E_{kin}$ will be $E_{kin}=m_q(\gamma-1)$ and we
can estimate $\Delta\gamma$ from the relation $\Delta
E_{kin}=m_q \cdot \Delta\gamma=2\cdot T_q \cdot v_{2}$.

\begin{figure}[thb]
\includegraphics[width=1.0\linewidth]{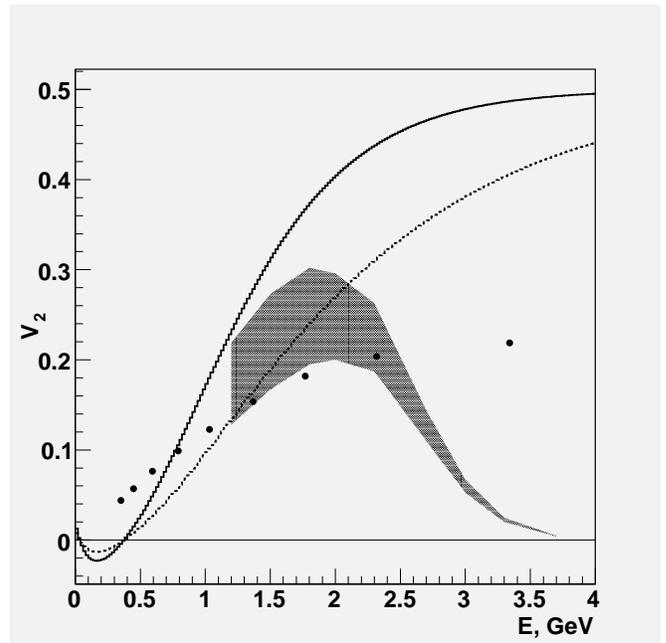}
\caption{\label{fig:v2} Fig.2. Azimuthal asymmetry parameter $v_2$ for thermal photons versus 
photon energy in 30-35\% centrality Au+Au collisions. Solid line -- 
for evaluation using the constituent quark mass $m_q=0.1$ GeV, dashed line
-- for $m_q=0.3$~GeV. 
Experimental points for non-identified hadrons
 in Au+Au collisions at centrality 30-40\% versus hadron transverse
 kinetic energy are from~\cite{ppg062}. 
Shaded area shows change of the result if we take into
account the contribution from direct photons.}
\end{figure}

At our maximum transverse boost value $\beta$ =0.75 (averaged over all angles versus the 
reaction plane), 
$\gamma$=1.51. If $T_q$=0.1 $GeV$, $v_2$=0.07  and taking a constituent quark mass $m=0.3$ $GeV$
we get $\Delta \gamma $=0.047. This gives estimates of the maximum 
velocity boost $\beta^{in}$=0.77 and $\beta^{out}$=0.73. 
Now we can estimate elliptic flow for thermal photons. 
Fig.~\ref{fig:v2} shows the value of $v_2$ for 30-35\% centrality Au+Au collisions 
assuming $m_q$=0.3~GeV (dashed 
line) versus photon 
energy for the case of $\beta_T$=0.75$c$, $T$=0.2 GeV and  
$\beta_L$=0.7$c$. We use the similar wounded nucleon 
distribution in the $x-y$ plane as before by calculating the photon spectrum. 
For comparison we plot the experimental data of $v_2$ for  
inclusive hadrons versus hadron transverse kinetic energy~\cite{ppg062}. 
We see significant 
value of $v_2$ for thermal photons, which is comparable with that for hadron
$v_2$ and then significantly overshoots hadron points at larger energy. We
have to take into account that in actual experiments, for photon energies above 2 $GeV$, there is a 
significant contribution from direct photons (thin line in
Fig.~\ref{fig:data}) for which $v_2$ is about zero~\cite{ppg046}, 
this dilutes the total photon $v_2$. We get an estimation of the relative direct photon yield from experimental 
results~\cite{data}. 
Thus, the measured value of photon $v_2$ could be 
surprisingly close (or even larger) to the hadron $v_2$. At low energy, where the spectrum
 is red shifted, $v_2$ gets negative.    
Selection of $m_q=0.3$~$GeV$ is quite arbitrary, so we also made
calculations for $m_q=0.07$~$GeV$. In this case we get $\Delta \gamma $=0.2, $\beta^{in}$=0.81
and $\beta^{out}$=0.65. 
As one can see, $v_2$ for thermal photons does not change much on the 
assumption of constituent quark mass in hadron. Very much the same results we get for
combinations T=0.17 GeV, $\beta_L$=0.5$c$, or T=0.15 GeV and $\beta_L$=0.

Resent
experimental data for $Pb$+$Pb$ collisions at LHC energy show large hadron $v_2$~\cite{lhc_v2}
with similar values as at RHIC. Thus, we can also expect
large and at about the same value of $v_2$ for thermal photons at LHC 
and red/blue shift effects as well. Similar features, seen for thermal photon spectrum, should also be 
observed for low invariant mass dilepton pairs.

In conclusion, based on a simple assumption that in nucleus-nucleus collisions most of the thermal photons 
are produced at the stage when the bulk flow is already developed,
we have examined the resulting change in the photon spectrum due to longituidinal and transverse expansion. The observed picture is very similar to the 
astrophysical expansion. 
Strong correlation was found between the temperature and the longitudinal flow parameters. 
In the absence of longitudinal flow, which is definitely not the case, we get the lowest estimate for 
the thermal photon temperature of about 0.15 $GeV$. We attribute the observed elliptic flow for hadrons 
to some amplitude modulation of the radial flow. From this we estimate the photon flow parameter $v_2$ 
for mid-central collisions, which are found to be large. Because of some contribution to the yield from thermal 
photons, which were produced before bulk flow developed, we can consider our estimates for v2 as upper limits. On the other hand, if in experiment we will see significant photon elliptic flow, it means 
that the most of photons are produced when radial and longitudinal flow are already developed. Unfortunately
we can not distinguish contributions from the quark-gluon or hadron gas stages. 

This work was partially supported by Russian Fund for Fundamental Investigations (RFFI) under grant 
number 08-02-00459-a. We would like to thank Roy Lacey for his help with grammar. 


\end{document}